\documentstyle[12pt]{article}
\def\ai{\'{\i}}

\def\e3o{e^{3\Omega}}

\def\-3o{e^{-3\Omega}}

\def\be{\begin{equation}}
\def\ee{\end{equation}}

\input{tcilatex}
\begin{document}

\baselineskip.33in

\centerline{\large{\bf DEPARAMETRIZATION AND QUANTIZATION}} % 
\centerline{\large{\bf OF THE TAUB UNIVERSE}} %
\centerline{\large{\bf }}

\vskip0.8cm

\centerline{GAST\'ON GIRIBET}

\centerline{\it  Instituto de Astronom\ai a y F\ai sica del Espacio} %
\centerline{\it C.C. 67, Sucursal 28 - 1428 Buenos Aires, Argentina} %
\centerline{\it E-mail: gaston@iafe.uba.ar }

\bigskip

\centerline{CLAUDIO SIMEONE}

\centerline{\it Departamento de F\'{\i }sica, Comisi\'{o}n Nacional de
Energ\'{\i }a At\'{o}mica} \centerline{\it Av. del Libertador 8250 - 1429
Buenos Aires, Argentina} \centerline{\it and} \centerline{\it Departamento
de F\'{\i }sica, Facultad de Ciencias Exactas y Naturales} \centerline{\it %
Universidad de Buenos Aires, Ciudad Universitaria} \centerline{\it %
Pabell\'{o}n I - 1428, Buenos Aires, Argentina} \centerline{\it E-mail:
simeone@tandar.cnea.gov.ar}

\vskip1cm

\noindent

ABSTRACT

\vskip1cm

Previous analysis about the deparametrization and path integral quantization of cosmological models are extended to models which do not admit an intrinsic time. The formal expression for the transition amplitude is written down for the Taub anisotropic universe with a clear notion of time. The relation existing between the deparametrization associated to gauge fixation required in the path
integral approach and the procedure of reduction of the Wheeler-De Witt equation is also studied.

\vskip1cm


\newpage

\vskip1cm

%%%%%%%%%%%%%%%%%%% CANONICAMENTE %%%%%%%%%%%%%%%

\section{Introduction}

%%%%%%%%%%%%%%%%% INTEGRAL DE CAMINOS %%%%%%%%%%%

A time dependent canonical transformation can turn a parametrized system into an ordinary gauge system. Indeed, in reference \cite{rafaclo} it has been shown how
       to build a canonical transformation such that the fixation of the new coordinates is equivalent to the fixation of the
       original ones. More recently, and within this context, simple cosmological models were used as examples to show that gravitation can be quantized as an ordinary gauge system if the
       Hamilton-Jacobi is separable. In this paper we discuss how to extend this procedure in order to study the case of Hamiltonian constraints which do not allow for an intrinsic global time \cite{hc}.
We treat the case of the Taub anisotropic universe as an example of physical interest with the intention to present a consistent deparametrization procedure which leads to write down a formal expression for the quantum transition amplitude. We study the realization of the analysis of the reduction
within the context of the path integral approach and we show that the
extrinsic time identified in \cite{gabirafa} coincides with one which can be
obtained by means of the systematic method proposed previously in references 
\cite{rafaclo}\cite{hc}. We will also discuss that this coincidence is, in
fact, an evidence of the narrow relation existing between the canonical
approach of quantization and the path integral quantization with a clear
notion of time. Then, we analyse aspects of the structure of the constraint
surface in order to establish the correspondence.

In section 2 we analyse the steps of deparametrization and path integral quantization of the Taub universe.
Then, in section 3, we discuss the canonical quantization, leaving the conclusions for
section 4.

\section{Path integral approach}

Let us consider a Hamiltonian constraint $H$ with the generic form 
\begin{equation}
H=G^{ij}\pi _{i}\pi _{j}+\sum_{n=0}^{N}c_{n}e^{\alpha _{i}^{n}g^{i}}\approx
0  \label{tiroloco}
\end{equation}
where $G^{ij}$ represents the minisuperspace reduction of a supermetric with
Lorentzian signature and \{$g^{i},\pi ^{i}$\} are the degrees of freedom of
the system and its conjugate momenta respectively. The indices $i,j\in
\{0,...d\}$ and $n\in \{0,...N\}$, being $N\leq d$.

In fact, it is possible to prove that\ if the matrix with components \{$%
\alpha _{i}^{n}$\} admits a dimensional extension \{$\alpha _{i}^{n}$\}$%
\rightarrow $\{$\alpha _{i}^{j}$\} leading to obtain a matrix $\alpha
_{i}^{j}$ of dimension $d+1$ that is diagonalizable in a base \{$x^{i}$\} in
which the form of the supermetric $G_{ij}$ is also diagonal, then a
canonical transformation exists leading to obtain a classically equivalent
Hamiltonian constraint which admits an intrinsic global phase time.

An example of this kind of constraint (\ref{tiroloco}) is given by the Taub
cosmological model, which represents a particular case of the anisotropic
generalization of the Friedmann-Robertson-Walker universe with curvature $%
k=+1$ ($i.e.$ by setting to zero one of the degrees of freedom of the
diagonal Bianchi type-IX universe). Other models of this class can be found
in reference \cite{salopec}. In this paper, we explicitly apply the
procedure described above to the case of the Taub universe in order to turn
this cosmological model into an ordinary gauge system and impose on it
canonical gauges to identify a global phase time.

The action functional of the Taub universe reads

\begin{equation}
S=\int_{\tau _{1}}^{\tau _{2}}\left( \pi _{+}{\frac{d\beta _{+}}{d\tau }}%
+\pi _{\Omega }{\frac{d\Omega }{d\tau }}-N{\cal H}\right) d\tau ,
\end{equation}
with the Hamiltonian constraint 
\begin{equation}
{\cal H}=e^{-3\Omega }(\pi _{+}^{2}-\pi _{\Omega }^{2})+{\frac{1}{3}}%
e^{\Omega }(e^{-8\beta _{+}}-4e^{-2\beta _{+}})\approx 0.
\end{equation}
Because the factor $e^{-3\Omega }$ is positive definite, this constraint is
equivalent to 
\begin{equation}
H=\pi _{+}^{2}-\pi _{\Omega }^{2}+{\frac{1}{3}}e^{4\Omega }(e^{-8\beta
_{+}}-4e^{-2\beta _{+}})\approx 0.  \label{h}
\end{equation}
Here, we identify the previous nomenclature with physical parameters as ($%
g^{0},g^{1}$)$\equiv $($\Omega ,\beta _{+}$).

Besides a separable Hamiltonian, the Taub cosmological model includes true
degrees of freedom and a potential which vanishes at a given point of the
phase space, so making impossible the definition of an intrinsic time in
terms of the original variables. As it is easy to see, for $\beta _{+}=-%
\frac{1}{6}\ln 4$ the potential is zero.

Note that in the case (\ref{h}) the parameters of the Hamiltonian are given by $\alpha
_{1}^{1}=\alpha _{1}^{2}=4,$ $\alpha _{2}^{1}=-8,$ $\alpha _{2}^{2}=-2$ and $%
c_{1}=-4c_{2}=\frac{1}{3}$. The Hamiltonian is not separable in terms of the
coordinates and momenta $(\Omega ,\beta _{+},\pi _{\Omega },\pi _{+})$.
Then, in order to apply the method of deparametrization we define the
coordinates 
\begin{equation}
x=\Omega -2\beta _{+},\ \ \ \ \ \ y=2\Omega -\beta _{+}  \label{variables}
\end{equation}
and, thus, we can write 
\begin{equation}
H=\pi _{x}^{2}-\pi _{y}^{2}+{\frac{1}{9}}(e^{4x}-4e^{2y})\approx 0.
\end{equation}

At this stage we then have $H=H_{1}(x,\pi _{x})+H_{2}(y,\pi _{y})$ with $%
H_{1}>0$ and $H_{2}<0$. Since the potential vanishes for $y=2x-\ln 2$, a
time in terms of only $x,y$ does not exist. Hence before turning the model
into an ordinary gauge system \cite{rafaclo}\cite{hc} we shall make a first
canonical transformation to the variables $(x,s,\pi _{x},\pi _{s})$ so that
the {\it new} potential has only one positive definite term. The resulting
new coordinates will correspond to a set $\{\tilde{q}^{i}\}$ in terms of
which an intrinsic time exists.

We shall perform a canonical transformation matching $H_{2}(y,\pi _{y})=-\pi
_{s}^{2},$ so that $\pi _{s}=\pm \sqrt{-H_{2}(y,\pi _{y})}$. This is
achieved \cite{gabirafa} by introducing the generating fuctions of the first
kind 
\begin{equation}
f_{1}(y,s)=\pm {\frac{2}{3}}e^{y}\sinh s.  \label{dieci}
\end{equation}
The momenta are then given by 
\begin{eqnarray}
\pi _{y} &=&\pm {\frac{2}{3}}e^{y}\sinh s  \nonumber \\
\pi _{s} &=&\pm {\frac{2}{3}}e^{y}\cosh s  \label{seis}
\end{eqnarray}
so that 
\[
\pi _{y}^{2}+{\frac{4}{9}}e^{2y}={\frac{4}{9}}e^{2y}(\sinh ^{2}s+1)=\pi
_{s}^{2} 
\]
and the Hamiltonian can be written as 
\begin{equation}
H(s,x,\pi _{s},\pi _{x})=-\pi _{s}^{2}+\pi _{x}^{2}+{\frac{1}{9}}%
e^{4x}\approx 0.  \label{9}
\end{equation}

It is important to note the reason why we have introduced two possible
definitions of $f_{1}$, indeed this ambiguity is reflected by the fact that
the constraint can be written as 
\begin{equation}
H=\left( -\pi _{s}+\sqrt{\pi _{x}^{2}+{\frac{1}{9}}e^{4x}}\right) \left( \pi
_{s}+\sqrt{\pi _{x}^{2}+{\frac{1}{9}}e^{4x}}\right) \approx 0,  \label{lobo}
\end{equation}
and if we had chosen a definite sign, according to the resulting sign of $%
\pi _{s}$, only one of the factors would be zero. But we shall consider the
Hamiltonian constraint (\ref{9}) as the starting point for applying the
deparametrization procedure, and we shall not go back to put the results in
terms of the original variables. Indeed, at the level of the variables $%
(x,s,\pi _{x},\pi _{s})$ there is no justification to prefer one possible
sign of $\pi _{s}$.

A point to be noted is the fact that the action $S[x,s,\pi _{x},\pi _{s},N]$
will differ from $S[\Omega ,\beta _{+},\pi _{\Omega },\pi _{+},N]$ in
surface terms associated to the transformation generated by $f_{1}$, namely $%
D$: 
\begin{eqnarray}
S[x,s,\pi _{x},\pi _{s},N] &=&\int_{\tau _{1}}^{\tau _{2}}\left( \pi _{x}{%
\frac{dx}{d\tau }}+\pi _{s}{\frac{ds}{d\tau }}-NH(x,s,\pi _{x},\pi
_{s})\right) d\tau  \label{actionc} \\
&=&S[\Omega ,\beta _{+},\pi _{\Omega },\pi _{+},N]+\left[ D(\tau )\right]
_{\tau _{1}}^{\tau _{2}},
\end{eqnarray}
and when we turn the system into an ordinary gauge one \cite{rafaclo} we
will have additional surface terms $B(\tau )$; but as we will not be
interested in a transition amplitude between states labeled by the original
coordinates $\{x,y\}$, but by $\{x,s\}$, we shall not require that $D(\tau
)+B(\tau )=0$, but only that $B(\tau )=0$ in a globally good gauge such that
an intrinsic time is defined \cite{rafaclo}. This ensures that the action of the ordinary gauge system weights the paths in the same way that the original parametrized action.

Now, let us introduce the coordinates 
\begin{equation}
u={\frac{1}{12}}e^{2(x+s)},\ \ \ \ \ \ \ \ \ v={\frac{1}{12}}e^{2(x-s)}
\label{a}
\end{equation}
which lead to the equivalent constraint 
\begin{equation}
H^{\prime }=\pi _{u}\pi _{v}+1.  \label{aa}
\end{equation}

After two succesive canonical transformations in order to obtain a system
with non conserved observables \cite{rafaclo}\cite{hc} we find the following
definition of variables of the gauge system, which arises from indentifying $%
P_{0}$ with the energy and going back to $(x,s,\pi _{x},\pi _{s})$: 
\begin{eqnarray}
Q^{0} &=&{\frac{e^{2(x-s)}}{12P}},  \nonumber \\
Q &=&{\frac{1}{12}}e^{2(x+s)}+{\frac{1}{P^{2}}}\left( {\frac{1}{12}}%
e^{2(x-s)}(1-P_{0})-\eta T(\tau )\right) ,  \nonumber \\
P_{0} &=&9(\pi _{x}^{2}-\pi _{s}^{2})e^{-4x}+1,  \nonumber \\
P &=&3(\pi _{s}+\pi _{x})e^{-2(x+s)}.
\end{eqnarray}
Here $T(\tau )$ is a monotonous function of $\tau $ and $\eta =\pm 1$. By
fixing the gauge $\chi \equiv Q^{0}-T(\tau )=0$ it is possible to see that
an extrinsic time is 
\begin{equation}
t(x,s,\pi _{x},\pi _{s})={\frac{1}{36}}{\frac{e^{4x}}{\pi _{x}+\pi _{s}}}.
\end{equation}

The constraint surface splits into two sheets given by the sign of $\pi _{s}$%
. On each sheet the intrinsic time can be defined as 
\begin{equation}
t(x,s)={\frac{1}{12}}sign(\pi _{s})e^{2(x-s)},
\end{equation}
which is associated to the canonical gauge $\chi \equiv Q^{0}P-\eta T(\tau
)=0$ because $P$ is proportional to $\pi _{s}+\pi _{x}$ and $sign(\pi
_{s}+\pi _{x})=sign(\pi _{s})$. The end point terms associated to the
transformation $(x,s,\pi _{x},\pi _{s})\to (Q^{i},P_{i})$ are \cite
{rafaclo} 
\begin{eqnarray}
B(\tau ) &=&2Q^{0}-Q^{0}P_{0}-2\eta {\frac{T(\tau )}{P}}  \nonumber \\
&=&{\frac{1}{3(\pi _{s}+\pi _{x})}}\left( {\frac{1}{6}}e^{4x}-2\eta
e^{2(x+s)}T(\tau )\right)
\end{eqnarray}
and with the gauge choice defining an intrinsic time they vanish on the
constraint surface $P_{0}=0$. The expression for the quantum propagator is 
\begin{equation}
\left\langle x_{2},s_{2}|x_{1},s_{1}\right\rangle =\int DQDP\exp \left[
i\int_{T_{1}}^{T_{2}}\left( PdQ-{\frac{\eta }{P}}dT\right) \right] .
\label{propagador}
\end{equation}

The end points are given as $T_{1}=\pm \frac{1}{12}e^{2(x_{1}-s_{1})}$ and $%
T_{2}=\pm \frac{1}{12}e^{2(x_{2}-s_{2})}$. Note that in the gauge defining
the intrinsic time the new coordinate $Q$ coincides with $\frac{1}{12}%
e^{2(x+s)}$, so that the paths go from $Q_{1}=\frac{1}{12}e^{2(x_{1}+s_{1})}$
to $Q_{2}=\frac{1}{12}e^{2(x_{2}+s_{2})}$. We have obtained a propagator
with a clear distinction between time and the physical degree of freedom.
The path integral corresponds to that for a conservative system with
Hamiltonian $\eta /P$. Because $\eta =sign(\pi _{s})$ then at level of the
physical degrees of freedom we have two disjoint theories, one for each
sheet of the constraint surface.

Propagators of the form (\ref{propagador}) have been previously treated
within the context of the minisuperspace approximation to quantization of
homogeneous cosmological models. In fact, the application of the
deparametrization procedure can be extended to several systems such as the
case of the Kantowski-Sachs model, the homogeneous Szekeres universe,
dilaton cosmological models and the closed de Sitter model (see \cite{libro}
and references therein for a complete review).

On the other hand, a global phase time can also be identified among the
coordinates labeling the quantum states. Because the momentum $\pi _{s}$
does not vanish on the constraint surface, the coordinate $s$ is itself a
time. More precisely, on each sheet of the constraint surface we can define an
intrinsic time in the form 
\begin{equation}
t^{*}\equiv -s~sign(\pi _{s}),  \label{tuno}
\end{equation}
because $\lbrack t^{*},H]=2\pi _{s} sign(\pi _{s})>0$. Note that this time can be directly recognized from eq. (\ref
{lobo}), as it was shown in reference \cite{gabirafa}.

Although we do not use this time as the time parameter in the path integral,
the interpretation of the result can be made more clear by recalling that
one of the coordinates which identify the states is a global phase time. In
fact, we can write the transition amplitude as 
\begin{equation}
\left\langle x_{2},t_{2}^{*}|x_{1},t_{1}^{*}\right\rangle .
\end{equation}

With this interpretation, the propagator becomes completely analogous to
that of a mechanical system with a physical degree of freedom $x$ whose
evolution is given in terms of a true time $t^{*}$.

At this point it is natural to ask wether this time could have been obtained
in a direct way with our deparametrization procedure. This is in fact
possible, even in the case that we include a matter field in the model.
Consider the Hamiltonian constraint for the Taub universe with a massless non interacting scalar field (a massless dust): 
\begin{equation}
H=-\pi _{\Omega }^{2}+\pi _{\phi }^{2}+\pi _{+}^{2}+{\frac{1}{3}}e^{4\Omega
}(e^{-8\beta _{+}}-4e^{-2\beta _{+}})\approx 0.
\end{equation}

If we change to the coordinates $x$ and $y$, and then perform the canonical
transformation generated by $f_{1}(y,s)$, we obtain the equivalent
constraint 
\begin{equation}
H=-\pi _{s}^{2}+\pi _{\phi }^{2}+\pi _{x}^{2}+{\frac{1}{9}}e^{4x}\approx 0,
\end{equation}
where we have redefined $\pi _{\phi }\to \pi _{\phi }/\sqrt{3}$. The
corresponding Hamilton--Jacobi equation is separable: 
\begin{equation}
-\left( {\frac{\partial W}{\partial s}}\right) ^{2}+\left( {\frac{\partial W%
}{\partial x}}\right) ^{2}+\left( {\frac{\partial W}{\partial \phi }}\right)
^{2}+{\frac{1}{9}}e^{4x}=E,
\end{equation}
and the solution is clearly of the form $W_{1}(x,\pi _{x})+W_{2}(\phi ,\pi
_{\phi })+W_{3}(s,\pi _{s})$. Introducing the integration constants $%
b^{2}=\pi _{\phi }^{2}$ and $a^{2}$ such that $a^{2}+b^{2}-E=\pi _{s}^{2}$
we obtain 
\begin{eqnarray}
W &=&sign(\pi _{x})\int dx\sqrt{a^{2}-{\frac{1}{9}}e^{4x}}  \nonumber \\
&&\mbox{}+s~sign(\pi _{s})\sqrt{a^{2}+b^{2}-E}+\phi ~sign(\pi _{\phi })\sqrt{%
b^{2}}.
\end{eqnarray}

If we match $E=\overline{P}_{0}$ then we have 
\[
\overline{Q}^{0}=\left[ {\frac{\partial W}{\partial \overline{P}_{0}}}%
\right] _{\overline{P}_{0}=0}=-sign(\pi _{s}){\frac{s}{2\sqrt{a^{2}+b^{2}}}}%
=-{\frac{s}{2\pi _{s}}}. 
\]
As $[\overline{Q}^{0},\overline{P}_{0}]=1$ then we can inmediately define an
extrinsic time as 
\begin{equation}
t(s,\pi _{s})\equiv \overline{Q}^{0}=-{\frac{s}{2\pi _{s}}}.
\end{equation}
As in the variables $(s,x,\phi ,\pi _{s},\pi _{x},\pi _{\phi })$ the
constraint surface is topologically equivalent to two disjoint half planes,
each one corresponding to $\pi _{s}>0$ and to $\pi _{s}<0$, we can define
the time as 
\begin{eqnarray}
t(s) &\equiv &2\pi _{s}sign(\pi _{s})\overline{Q}^{0}  \nonumber \\
&=&-s~sign(\pi _{s}),  \label{tdos}
\end{eqnarray}
which coincides with the time $t^{*}$ found just before following the steps
of the reduction proposed in reference \cite{gabirafa}. In terms of the new
variables this is an intrinsic time; of course, in terms of the original
variables it is an extrinsic time.

All the results include the sign of the momentum $\pi _{s}$, which comes
from the double sign in the definition of $f_{1}$; this fact manifestly
shows which is the counterpart of the ambiguous sign in the definition (\ref
{seis}) within the context of the path integral approach. However, some
authors have suggested that the constraint of a parametrized system, which
is linear in the momentum conjugated to the time, may be hidden in the
Hamiltonian formalism for the gravitational field.
According to this point of view, one should choose only one of both possible
signs for the generator $f_{1}$, and there would not be two coexisting
theories (see \cite{casta} for an interesting discussion about this point).
Our results, instead, reflect that we consider the quadratic Hamiltonian $%
H(x,s,\pi _{x},\pi _{s})$ as the starting point because our formalism
requires a constraint which, with a given choice of variables, admits an
intrinsic time, and at the level of this Hamiltonian there is no reason to
prefer one definite sign for the non vanishing momentum $\pi _{s}$. Anyway,
it must be signaled that, because the reduced Hamiltonian $\eta /P$ is
conserved, and in particular it does not vanish, in our interpretation there
are no transitions from states on one sheet to states on the other sheet of
the constraint surface, and therefore both points of view do not lead to an
essential contradiction.
\newpage
\section{\protect\smallskip Canonical quantization}

\subsection{Standard procedure}

In the literature we can find different solutions for the Taub universe. An
interesting example within those which do not start from an explicit
deparametrization is the solution found by Moncrief and Ryan \cite{mory} in
the context of an analysis of the Bianchi type-IX universe with a rather
general factor ordering of the Hamiltonian constraint \cite{haha}. In the
case of the most trivial ordering they solved the Wheeler--DeWitt equation
to obtain a wave function of the form 
\begin{equation}
\Psi (\omega ,\beta _{+})=\int_{0}^{\infty }d\omega F(\omega )K_{i\omega
}\left( {\frac{1}{6}}e^{2\Omega -4\beta _{+}}\right) K_{2i\omega }\left( {%
\frac{2}{3}}e^{2\Omega -\beta _{+}}\right) .
\end{equation}
where $K$ are the modified Bessel functions. In the particular case that $%
F(\omega )=\omega \sinh (\pi \omega )$ they have shown that the wave
function can be written in the form 
\begin{equation}
\Psi (\Omega ,\beta _{+})=R(\Omega ,\beta _{+})e^{-S}
\end{equation}
with $S$ a combination of exponentials in $\Omega $ and $\beta _{+}$. An
important feature of this wave function is that for values of $\Omega $ near
the singularity (that is, the scale factor near zero) the probability is
spread over all possible degrees of anisotropy given by $\beta _{+}$, while
for large values of the scale factor the probability is peaked around the
isotropic Friedmann--Robertson--Walker closed universe.

\subsection{Schr\"{o}dinger equation and boundary conditions}

A rather different approach beginning with the identification of a global
phase time can be found in \cite{gabirafa}, where the authors obtain a
Schr\"{o}dinger equation and its solutions are used to select a set of
solutions of the Wheeler--DeWitt equation. They start from a Hamiltonian
like that given in (\ref{h}), and solve the Wheeler--DeWitt equation 
\begin{equation}
\left( {\frac{\partial ^{2}}{\partial x^{2}}}-{\frac{\partial ^{2}}{\partial
y^{2}}}-{\frac{1}{9}}e^{4x}+{\frac{4}{9}}e^{2y}\right) \Psi (x,y)=0
\label{daniel}
\end{equation}
like Moncrief and Ryan to obtain a set of solutions of the form 
\begin{eqnarray}
\varphi _{\omega }(x,y) &=&\left[ a(\omega )I_{i\omega }(\frac{2}{3}%
e^{y})+b(\omega )K_{i\omega }(\frac{2}{3}e^{y})\right]  \nonumber \\
&&\times \left[ c(\omega )I_{i\omega /2}(\frac{1}{6}e^{2x})+d(\omega
)K_{i\omega /2}(\frac{1}{6}e^{2x})\right] ,  \nonumber \\
&&
\end{eqnarray}
with $I$ and $K$ the modified Bessel functions. Then they consider a change
of variables analogous to (\ref{variables}) but with only the minus sign, so
that the momentum $\pi _{s}$ is negative definite, and the time is then $t=s$%
; hence in (\ref{dieci}) the first factor is positive definite, and the
second one is a constraint linear in $\pi _{s}=\pi _{t}$ which leads to the
Schr\"{o}dinger equation 
\begin{equation}
\left( \pm i{\frac{\partial \Psi }{\partial t}}+\left( {\frac{1}{9}}e^{4x}-{%
\frac{\partial ^{2}}{\partial x^{2}}}\right) ^{\frac{1}{2}}\right) \Psi
(t,x)=0  \label{lopez}
\end{equation}

According to this interpretation, the contribution of the modified functions 
$I_{i\omega /2}(\frac{1}{6}e^{2x})$ is discarded, because they diverge in
the classicaly forbidden region associated to the exponential potential ${%
\frac{1}{9}}e^{4x}$; the functions $I_{i\omega }(\frac{2}{3}e^{y})$,
instead, are not discarded, because in this picture the coordinate $y$ is
associated to the definition of time.

As it was mentioned before, an interesting point to be noted is the fact that
this time $t$ coincides with the definition (\ref{tdos}), including the
ambiguity in its sign which is related with the existence of two sheets of
the constraint surface.

A point that should be remarked is the fact that the utilization of the
Schr\"{o}dinger equation in order to select the physical solutions of the
Wheeler-De Witt equation has been possible because the potential in the
square root does not depend on time. This ensures that the constraints (\ref
{daniel}) and (\ref{lopez}), which are equivalent at the classical level,
are also equivalent in their quantum version as no additional terms arise
associated with conmutators \cite{futuro}.

\subsection{Wheeler--DeWitt equation with extrinsic time}

Consequently with the above remarks we propose to solve a Wheeler-De Witt equation starting from a form of the Hamiltonian constraint such that a global phase time is easily identified among the canonical coordinates; hence, the resulting wave function has an evolutionary form and it could be understood as it is in
ordinary quantum mechanics.

The constraint (\ref{h}) allows to trivially define the time as 
\begin{equation}
t=-s\,sign(\pi _{s}).
\end{equation}

We can make the usual substitution $p_{k}\to -i\frac{\partial }{\partial
q^{k}}$ to obtain the Wheeler--DeWitt equation 
\begin{equation}
\left( {\frac{\partial ^{2}}{\partial x^{2}}}-{\frac{\partial ^{2}}{\partial
s^{2}}}-{\frac{1}{9}}e^{4x}\right) \Psi (x,s)=0.
\end{equation}
This equation admits the set of solutions 
\begin{eqnarray}
\Psi _{\omega }(x,s) &=&\left[ a(\omega )e^{i\omega s}+b(\omega )e^{-i\omega
s}\right]  \nonumber \\
&&\times \left[ c(\omega )I_{i\omega /2}(\frac{1}{6}e^{2x})+d(\omega
)K_{i\omega /2}(\frac{1}{6}e^{2x})\right] ,  \nonumber \\
&&
\end{eqnarray}
where $\pm s$ is a global phase time.

There are some points which deserve certain analysis: The first one is that
we have chosen the real values for $\omega $; the reason for this choice is
the requirement to obtain the free particle solutions in the asymptotically
free regime. The second point is that we have not discarded negative energy
solutions; this would have been equivalent to select one sheet of the
constraint surface. As we pointed in the context of path integral
quantization, once we decide to work and give the results in terms of the
variables $(x,s,\pi _{x},\pi _{s})$ there is no reason to choose one sign
for the momentum $\pi _{s}$. Moreover, if we take the quadratic form of the
constraint as an essential feature of gravitation, we should only admit a
canonical transformation leading to a constraint equivalent to the original
one; hence, both signs of $f_{1}$ must be considered simultaneously to
ensure that the original constraint and that in terms of the new coordinates
and momenta yield differential equations with the same number of solutions.
One half of the solutions of our Wheeler--DeWitt equation correspond to
those of the preceding section, which yielded from a Schr\"{o}dinger
equation. An important fact is that our procedure allows to obtain them
without the necessity of defining a prescription for the square root
operator, but only by choosing the trivial factor ordering. Finally, we should stress that although the original momenta cannot be avoided in the description, their role is restricted to appearing in the time variable, but not in the physical degree of freedom $x$.

\subsection{Another extrinsic time for the Taub universe}

On the other hand, in the case of the Taub universe a time in terms of the
original variables can be found in a straightforward way by identifying the
coordinate $\overline{Q}^{0}$ conjugated to $\overline{P}_{0}\equiv H$. The
Hamilton--Jacobi equation associated to the constraint $H$ is 
\begin{equation}
3\left( {\frac{\partial W}{\partial x}}\right) ^{2}-3\left( {\frac{\partial W%
}{\partial y}}\right) ^{2}+{\frac{1}{3}}e^{4x}-{\frac{4}{3}}e^{2y}=E.
\end{equation}
The solution is clearly of the form 
\begin{equation}
W=W_{1}(x)+W_{2}(y)
\end{equation}
where 
\[
W_{1}(x)=\pm {\frac{1}{3}}\int \sqrt{3\alpha ^{2}-e^{4x}}dx 
\]
with $\pm =sign(\pi _{x})$, and 
\[
W_{2}(y)=\pm {\frac{1}{3}}\int \sqrt{3(\alpha ^{2}-E)-4e^{2y}}dy 
\]
with $\pm =sign(\pi _{y})$. Matching the constants $\alpha $ and $E$ to the
new momenta $\overline{P}$ and $\overline{P}_{0}$ and following the
procedure of Chapter 3 we have 
\begin{eqnarray}
\overline{Q}^{0} &=&\left[ {\frac{\partial }{\partial \overline{P}_{0}}}%
\left( \pm {\frac{1}{3}}\int \sqrt{3(\overline{P}^{2}-\overline{P}%
_{0})-4e^{2y}}dy\right) \right] _{\overline{P}_{0}=0}  \nonumber \\
&=&\mp {\frac{1}{4}}{\frac{1}{\sqrt{3\overline{P}^{2}}}}\ln \left( {\frac{%
\sqrt{3\overline{P}^{2}}-\sqrt{3\overline{P}^{2}-4e^{2y}}}{\sqrt{3\overline{P%
}^{2}}+\sqrt{3\overline{P}^{2}-4e^{2y}}}}\right)
\end{eqnarray}
with $-$ for $\pi _{y}>0$ and $+$ for $\pi _{y}<0$. Because on the
constraint surface $\overline{P}_{0}=0$ we have $3\pi
_{y}^{2}+(4/3)e^{2y}=\alpha ^{2}=\overline{P}^{2}$ then $\pi _{y}=\pm {\frac{%
1}{3}}\sqrt{3\overline{P}^{2}-4e^{2y}}$ and hence for both $\pi _{y}>0$ and $%
\pi _{y}<0$ we obtain 
\begin{equation}
\overline{Q}^{0}={\frac{1}{4\sqrt{9\pi _{y}^{2}+4e^{2y}}}}\ln \left( {\frac{%
\sqrt{9\pi _{y}^{2}+4e^{2y}}+3\pi _{y}}{\sqrt{9\pi _{y}^{2}+4e^{2y}}-3\pi
_{y}}}\right) .
\end{equation}

The gauge can be fixed by means of the canonical condition $\chi \equiv 
\overline{Q}^{0}-T(\tau )=$ with $T$ a monotonic function. Thus, as $3\pi
_{y}^{2}+\frac{4}{3}e^{2y}=\alpha ^{2}>0$, we can define an extrinsic time
as 
\begin{eqnarray}
t(\pi _{y}) &\equiv &12|\alpha |\overline{Q}^{0}  \nonumber \\
&=&\ln \left( {\frac{\sqrt{3\alpha ^{2}}+3\pi _{y}}{\sqrt{3\alpha ^{2}}-3\pi
_{y}}}\right) .
\end{eqnarray}
Now, if we go back to the original variables $(\Omega ,\beta _{+},\pi
_{\Omega },\pi _{+})$ we obtain 
\begin{equation}
t(\pi _{\Omega },\pi _{+})=\ln \left( {\frac{\sqrt{3\alpha ^{2}}-(\pi
_{\Omega }+\pi _{+})}{\sqrt{3\alpha ^{2}}+(\pi _{\Omega }+\pi _{+})}}\right)
.
\end{equation}

Again, it could be interesting to notice that the existence of different
well defined global extrinsic times confirms that the coincidence of the
time (\ref{tuno}) and (\ref{tdos}) shows the close relation existing between
the path integral approach and the analysis of the reduction proposed in
reference \cite{gabirafa}.

\section{Discussion}

The requirement of definition of a global phase time in terms of the momenta
could suggest that we should abandon the idea of obtaining an amplitude for
states characterized by the original coordinates. However, while a
deparametrization in terms of the momenta may be completely valid at the
classical level, it has been pointed by Barvinsky that at the quantum level
there is an obstacle which is peculiar of gravitation: There are basically
two representations for quantum operators, the coordinate representation and
the representation in which the states are characterized by occupation
numbers associated to given values of the momenta. The last one is
appropriate when the theory under consideration allows for the existence of
assimptotically free states associated to an adiabatic vanishing of
interactions, so that a natural one-particle interpretation in terms of
creation and annihilation operators exists. In quantum cosmology these
assymptotic states do not, in general, exist. The suitable representation
must be able to handle with essentially non linear and non polynomial
interactions, and such a representation is a coordinate one. In this representation the quantum states are represented by wave functions in terms of the
coordinates. The usual Dirac--Wheeler--DeWitt quantization with momentum
operators in the coordinate representation acting on $\Psi (q)$ follows this
line; but, as we have already observed, this formalism is devoided of a
clear notion of time and evolution, unless the potential is everywhere non
null so that we can find a time among the canonical coordinates.

Within this context, it would be interesting to explore the canonical
transformations which lead to classical systems with a well defined
intrinsic time as a systematic procedure of deparametrization. In fact, this
it is one of the central aspects of the analysis presented in \cite{gabirafa}%
, where the canonical transformation (\ref{dieci}) allows to obtain a system
for which the Hamiltonian constraint admits the possibility to define an
intrinsic time ({\it i.e.} in terms of the {\it new} variables). This scheme
of work is justified by the observation above.

Taking into account the importance of this proposal, here we have discussed
the relation between the mechanism of reduction of the Taub cosmological
model proposed by Catren and Ferraro and the path integral approach, finding
that both lead to the identification of the same extrinsic time in terms of
the original variables. We have also commented the way in which the
ambiguity in the election of the constraint sheet manifestly appears in both
frameworks. On the other hand, a secondary result of our work is the fact
that this is the first application of the path integral approach presented
in \cite{hc} to the Bianchi universes.

\[
\]

{\bf Acknowledgements:} The authors are grateful to Esteban Calzetta for
useful conversations. This work was supported by CNEA and CONICET.

\end{document}